\documentclass{ws-p9-75x6-50}

\begin{document}
\title{Is a Coherent Picture of Massive Neutrinos Emerging?}
\author{Francesco Vissani}
\address{INFN, Lab.\ Naz.\ Gran Sasso, 
Theory Group, I-67010 Assergi (AQ), Italy\\E-mail: vissani@lngs.infn.it}


\maketitle

\abstracts{The hypothesis
that neutrinos are massive has a strong 
experimental support; however, 
the information we have is quite 
limited, and many possibilities are open.
Theoretical considerations might help 
fill the gaps (or foresee regularities), 
as we illustrate with a specific example.}

Several experiments suggest that the 
neutrinos oscillate in flavor, as expected if they are massive;
the strongest indication is the atmospheric neutrino deficit. 
However, most of available information 
can be dubbed as flavor {disappearance}
and this is perhaps the main reason of doubt
(the signal of $\bar{\nu}_{\rm e}$ appearance 
seen in the LSND experiment will be 
tested soon).
More specific statements rely on
details of the data or analyses, and have
a less compelling significance at present\footnote{SuperKamiokande (SK)
obtained neutral currents enriched samples, becoming sensitive to 
$\nu_\tau$ oscillated from atmospheric $\nu_\mu;$
and the quality of $L/E$ tests is improving (SK, MACRO, Soudan2).
Since $\sigma({\rm \nu_\mu e})/ \sigma({\rm \nu_e e})\approx 1/6$
at $E_\nu\sim 7$ MeV, SK has also some sensitivity 
to an oscillated flavor for solar neutrinos; 
SNO can improve on that, using
 $\nu D\to \nu\ p\ n$ ($\nu=\nu_{\rm e,\mu,\tau}$).}.

The safest approach would be to wait for more data.
If instead one tries to interpret all the
indications by oscillations, one has to 
answer the Sphinx question: Are three neutrinos enough?
If the negative opinion is expressed, one has to confront with
sterile neutrinos, 6 possible arrangements
of the mass levels, not to speak of the 
number of mixings and phases.
Suspending the judgement on LSND indications, 
we can instead explain the deficit of atmospheric and solar  
neutrinos with \underline{three neutrinos only}.
Since we just need to have two different 
frequencies of oscillation (=doublets of levels)
there are 2 spectra compatible with what we know.
One of these spectra looks very different from 
those of charged fermions (``inverted'' 
spectrum).  Both of them, 
in principle, could be offset by a common 
mass scale, that does not affect oscillations (up 
to the case of quasi ``degenerate'' spectra).
The best bet possibility seems to be the one 
where the spectrum looks ``charged-fermion like'';
\underline{``normal'' spectrum, no large mass offset}.
We take this case as reference, and concentrate 
the discussion on 4 questions:
{\em Q1\ What is the weight of neutrinos? 
Q2\ Are neutrinos Majorana particles? 
Q3\ Are large angles maximal? 
Q4\ Do oscillation with ``atmospheric'' frequency
involve all flavors?}
(to rephrase the latter one; 
Do we have plain $\nu_\mu\to \nu_\tau$ oscillations, 
or also a bit of $\nu_{\rm e}?$).
In comparison, the question in the title seems 
easy--the answer being a plain ``no''.

So, we try with some theory.
Since we have been lead to assume that the largest
mass scale is $(\Delta m^2_{atm})^{1/2}=40-70$ meV,
no doubt that {\em Q1} is hard for experiments.
Now; how does the mass matrix looks keeping {\em only} 
this scale? Here is the answer:
\begin{equation}
{\cal M}= m_3\cdot v_3^t \otimes v_3, 
         \mbox{where }v_3\approx (\varepsilon,s,s)
          \mbox{ and } m_3 = (\Delta m^2_{atm})^{1/2};
\label{eq1}
\end{equation}
the approximate equality of the $\nu_\mu$ and $\nu_\tau$
components accounts for {almost maximal} 
{atmospheric mixing}
($s^{-1}=\sqrt{2}$---with {\em Q3} called into 
play by the sign ``$\approx$'');
{\em Q4} can be reformulated as: What is the 
value of $\varepsilon$ 
(the upper bound being $\sim 1/6$). 
Carrying on the outer product $\otimes$, one
notes that the $\nu_\mu-\nu_\tau$ sector
is the ``\underline{dominant block}'' of the mass matrix 
(with null determinant),
whereas the entry ${\cal M}_{\rm ee}$ arises 
at order $\varepsilon^2$ (that is sad if we want to answer 
{\em Q2} experimentally\footnote{A crucial difference
between {\em Q1} and {\em Q2} lies in the foreseen 
sensitivity of the experimental setups; more precisely,
in the alternative hypothesis ${\cal M}_{\rm ee}\sim m_3,$ 
realistic neutrinoless-double beta decay experiments 
would find a signal, whereas $m_{\nu_{\rm e}}\sim m_3$ 
would be invisible in tritium endpoint spectra.}).
What is the effect of the other two neutrinos? 
The relevant mass scale is 
$m_{sol}=(\Delta m^2_{sol})^{1/2}<15$ meV, and the range
$5-9$ meV is favored by present solar neutrino data
(with Homestake results playing an important role). 
As a reference value, 
we will keep in mind the
mass ratio $m_{sol}=m_{atm}/10.$ 
We add now to the mass matrix
the terms $m_2\cdot v_2^t \otimes v_2$ and  $m_1\cdot v_1^t \otimes v_1,$  
with $m_2^2-m_1^2=m_{sol}^2$ (and $m_1$ not much larger than
$m_{sol}$ in view of the discussion above) 
with the condition that the 3 versors $v_i$ are orthogonal. 
Since $v_3$ describes the state
$\nu_3\sim \nu_{+}\equiv (\nu_\mu+\nu_\tau)/\sqrt{2},$ 
$v_{1,2}$ describe the system\footnote{Account of additional
terms order $\varepsilon$ changes the mass matrix by 
very little amount.}   
$\nu_{\rm e}-\nu_{-};$
the solar neutrino deficit 
requires them to be mixed. 
The main consequences of 
the new terms are that: {\em a)} the element ${\cal M}_{\rm ee}$ 
seems to receive a contribution order $m_3/10$ 
(that might be sufficient to address {\em Q2} experimentally);
{\em b)} the elements ${\cal M}_{\rm e\mu}$ and ${\cal M}_{\rm e\tau}$ 
become different, but presumably remaining 1/10 
of the larger ones or so;
{\em c)} the other elements get minor 
corrections, that however turn the determinant 
of the dominant block non-zero. 
One can force a bit the matter and assume that the 
{\em structure} of the mass matrix ${\cal M}$
(namely, the powers of $\varepsilon$)
remain almost the same, up to unknown factors ${\cal O}$(1).
Indeed, the parameter 
{${\cal M}_{\rm ee}$ {\em could} remain order $\epsilon^2,$}
if the contributions from the 2 lighter
neutrinos have opposite sign and tend to compensate each other. 
We hit then:
\begin{equation}
{\cal M}\stackrel{{\cal O}(1)}{=}
\frac{\langle H\rangle^2}{M_X}
\cdot
\left(\begin{array}{ccc}
\epsilon^2 & \epsilon & \epsilon \\
\epsilon &  1 & 1 \\
\epsilon &  1 & 1 
\end{array}\right)\ \ \ 
\mbox{with }\left\{\begin{array}{l}
\langle H\rangle=174\mbox{ GeV}\\
M_X=(8-16)\cdot 10^{15}\mbox{ GeV}
\end{array}\right.
\label{eq2}
\end{equation}
with $\varepsilon\sim m_\mu/m_\tau$ 
suggested by charged lepton masses and U(1) family symmetry. 
Grand unification seems to underlie the neutrino mass scale.
Now let us go from eq.\ \ref{eq2} onward.
A little determinant in the dominant block 
comes from unknown ${\cal O}(1)$ coefficients 
with a little but not unreasonable chance.
In this respect, the \underline{triplet} 
mechanism for neutrino mass is 
more predictive than usual seesaw;
it disfavors large hierarchies, favoring thus 
the solar neutrino solutions with large $m_{sol}.$

{\em Q1} and {\em Q2} would be too difficult. 
{\em Q3} would get negative answers (but model 
does not give its best for solar neutrinos). 
The answer to {\em Q4} would be yes;
the model would make a point if the 
electronic mixing $U_{\rm e3}$ is $\sim 0.04$ 
($\approx\varepsilon,$ up to ${\cal O}(1)$ factors).
To summarize: The picture we discussed 
encourages other theoretical efforts, and leads us to 
expect that the future will be again in oscillations 
experiments.


\begin{thebibliography}{99}

\bibitem{refs} The present study developed from
J Sato and T Yanagida, \Journal{\PLB}{430}{127}{1998}, 
{\em Nucl. Phys. Proc. Suppl.} {\bf 77}, 293 (1999),
\Journal{\PLB}{493}{356}{2000};
F Vissani, {\em JHEP} {\bf 9811}, 025 (1998);
M Tanimoto, {\tt hep-ph/0010088}.

\end{thebibliography}
\end{document}